\documentclass[twocolumn,showpacs,preprintnumbers,amsmath,amssymb]{revtex4}

\usepackage{graphicx}
\usepackage{dcolumn}
\usepackage{bm}
\def\overlinen#1{{\bar{#1}}}

\def\calS{{\mathcal{A}}}

\def\|{|\!|}

\newcommand{\bea}{\begin{eqnarray}}\newcommand{\eea}{\end{eqnarray}}
\newcommand{\ba}{\begin{array}}\newcommand{\ea}{\end{array}}
\newcommand{\bit}{\begin{itemize}}\newcommand{\eit}{\end{itemize}}
\newcommand{\ben}{\begin{enumerate}}\newcommand{\een}{\end{enumerate}}

\newcommand{\lf}{\left}

\newcommand{\noi}{\noindent}\newcommand{\non}{\nonumber}

\newcommand{\ran}{\rangle}
\newcommand{\ri}{\right}

\newcommand{\al}{\alpha}
\newcommand{\bt}{\beta}\newcommand{\ga}{\gamma}

\newcommand{\De}{\Delta}

\usepackage{color}
\usepackage{graphics}
\usepackage{epsf}
\newcommand{\bs}{\begin{slide}}
\newcommand{\es}{\end{slide}}
\def\boll#1{\mbox{\boldmath\footnotesize $#1$\normalsize\unboldmath}}
\def\bol#1{\mbox{\boldmath $#1$\unboldmath}}

\begin{document}

\title{Quantum behavior of deterministic systems with information loss:\\
Path integral approach }\author{M. Blasone}
\email{blasone@sa.infn.it}
\author{P. Jizba}
\email{p.jizba@fjfi.cvut.cz}
\author{H. Kleinert}
\email{kleinert@physik.fu-berlin.de} \affiliation{
${}^{*}$Dipartimento di  Fisica, Universit\`{a} di Salerno, I-84100 Salerno, Italy\\
${}^{\dag}$FNSPE, Czech Technical University, B\v{r}ehov\'{a} 7, 115
19 Praha 1, Czech Republic\\
${}^{\ddag}$Institut f\"{u}r Theoretische Physik, Freie
Universit\"{a}t Berlin, Arnimallee 14 D-14195 Berlin, Germany}

\begin{abstract}
't~Hooft's derivation of quantum from classical physics is analyzed
by means of the classical path integral of Gozzi {\em et al.}. It is
shown how the key element of this procedure -- the loss of
information constraint --- can be implemented by means of
Faddeev-Jackiw's treatment of constrained systems. It is argued that
the emergent quantum systems are identical with systems obtained in
[Phys.Rev. A{\bf71} (2005) 052507] through Dirac-Bergmann's
analysis. We illustrate our approach with two simple examples --
free particle and linear harmonic oscillator. Potential Liouville
anomalies are shown to be absent.
\end{abstract}

 \maketitle

\section{Introduction}

The idea of quantum mechanics as  the low-energy limit of some
more fundamental deterministic dynamics~\cite{monographs,elzeII}
has been revived recently by G.'t Hooft~\cite{tHooft,tHooft22}, in
the attempt for a radical solution of the so-called holographic
paradox, originally formulated in the context of black-hole
thermodynamics ~\cite{tHooft2,Bousso}.

There is a widespread negative attitude towards the possibility
of deriving quantum from classical physics which relies on Bell's
inequalities~\cite{bell}. However, although being clear that
quantum mechanics at laboratory scales violates these inequalities, a
common prejudice
is  that Bell's theorem should be true at all scales. As observed by
't Hooft~\cite{tHooft}, this need not be the case because such
fundamental concepts as rotational symmetry, isospin or even
Poincar\'{e} invariance
--- on which the usual forms of the Bell inequalities are based
--- may simply cease to exist at
Planck scale.

By resorting to simple dynamical systems, 't~Hooft has shown that
an appropriate constraining procedure applied to the
deterministic system, can reduce the physical
degrees of freedom so that quantum mechanics emerges. Such a
reduction of the degrees of freedom may be physically implemented by
a mechanism of information loss (dissipation).
This idea has been further developed by several authors
\cite{BJK,tHooft22,BJV3,Elze,Halliwell:2000mv,tHooft3,BJV1}, and
it forms the basis also of this paper.

Our aim is to study 't~Hooft's quantization procedure by means of
path integrals, along the line of what done
in our previous work~\cite{BJK}.  However,
in contrast to Ref.~\cite{BJK} here we treat 't~Hooft's constrained dynamics
 by means
of the Faddeev-Jackiw technique~\cite{F-J}. The constrained dynamics
enters into 't~Hooft's scheme twice: first, in the classical
starting Hamiltonian which is of first order in the momenta and thus
singular in the Dirac-Bergmann sense~\cite{Dir2}. Second, in the
information loss condition that one has to enforce in order to
achieve quantization~\cite{BJK}.  In our previous paper~\cite{BJK}
we have adopted the customary Dirac-Bergmann technique, which is
often cumbersome. Here, we want to point out the simplifications
arising from the alternative Faddeev-Jackiw method, which turns out
to admit a clearer exposition of the basic concepts.

The paper is organized as follows: In Section~II, we briefly discuss
the main features of 't\,Hooft's scheme. By utilizing the
Faddeev-Jackiw procedure we present in Section~III  a Lagrangian
formulation of 't\,Hooft's system, which allows us to quantize
't\,Hooft's system via path integrals in configuration space. It is
shown that the fluctuating system produces a classical partition
function. In Section IV, we make contact with Gozzi's superspace
path integral formulation of classical mechanics. In Section V, we
introduce 't\,Hooft's constraint which accounts for information
loss. This is again handled by means of Faddeev-Jackiw analysis.
Central to this analysis is the fact that 't~Hooft's condition
breaks the BRST symmetry and allows to recast the classical
generating functional into a form representing a genuine
quantum-mechanical partition function. In Section~VI, we present two
simple applications of our formalism. Associated technical details
of the anomaly cancelation are relegated to Appendix~A. A final
discussion is given in Section~VII.

\section{'t Hooft's quantization procedure}
In this section we  briefly review the main aspects of 't Hooft's
quantization procedure~\cite{tHooft22,tHooft3} to be used in this
work. The basic idea is  that there exists a simple class of
classical  systems that can be described by means of Hilbert space
techniques without loosing their deterministic character.
Only after enforcing certain constraints expressing information loss,
one obtains {\em bona fide} quantum systems.
Thus, the quantum states of actually observed degrees of
freedom ({\em observables}) can be identified with equivalence
classes of states that span the original (primordial) Hilbert
space of truly existing degrees of freedom ({\em be-ables}).

Such a scheme is realized in certain model quantum cases where one
may indeed identify the primordial systems of {\em be-ables} that
are entirely deterministic. In discrete-time systems this scenario
has been successfully applied, e.g., to cellular automata with
embedded information loss~\cite{tHooft} where the equivalence
classes were invoked to obtain a unitary evolution operator with a
genuine quantum mechanical Hamiltonian. Further examples of
discrete-time systems can be found, e.g., in Refs.~\cite{BJV3,BJV1}.

In the continuous cases the equivalence classes are tightly linked
with the loss of information condition --- that is represented by
a suitably chosen first-class primary constraint -- and ensuing
gauge freedom. As only the continuous times will be of concern
here let us briefly address the key points thereof.
We begin by observing that classical systems of the form
\begin{eqnarray}
 H({\bol{p}}, {\bol{q}}) =f^{a}({\bol{q}})p_{a}\, , \label{2.B.1} \end{eqnarray}
with repeated indices summed, evolve deterministically even after
quantization \cite{tHooft3}. This happens since in the Hamiltonian
equations of motion
\begin{eqnarray} \dot{q}^a &=& \{q^{a}, H\} = f^{a}({\bol{q}})\, , \label{2.B.2a} \\ [3mm]
\dot{p}_{a} &=& \{p_{a}, H\} = - p_{a}
\frac{\partial{f^{a}({\bol{q}})}}{\partial q^{a}}\, , \label{2.B.2b}
\end{eqnarray}
the equation for the $q^a$  does not contain $p_a$, making the
$q^a$ {\em be-ables}.
%
Because of the autonomous character of the dynamical equations
(\ref{2.B.2a}) we can always decide to define a formal Hilbert
space spanned by the states $\{| {\bol{q}} \rangle \}$, and define
the associated momenta $\hat{p}_a = - i \partial/ \partial q^a$.
The quantum mechanical ``Hamiltonian" generating (\ref{2.B.2a}) is
then $\hat{H} = f^a(\hat{{\bol{q}}})\hat{p}_a$. Indeed, due to
linearity of $\hat{H}$ in $\hat{p}_a$ we have that $\hat{q}^a(t +
\Delta t) = F^a[\hat{\bol{q}}(t), \Delta t]$ ($F^a$ is some
function) and hence $[\hat{q}^a(t), \hat{q}^b(t')] = 0$ for any
$t$ and $t'$. This in turn implies that the Heisenberg equation of
the motion for $\hat{q}^a(t)$ in the ${\bol{q}}$-representation is
identical with the $c$-number dynamical equation (\ref{2.B.2a}).

The basic physical problem with systems described by the Hamiltonian
(\ref{2.B.1}) is that they are not bounded from below. This defect
can be repaired in the following way \cite{tHooft3}: Let
$\rho(\hat{\bol{q}})$ be some positive function of $\hat{q}_a$ with
$ [ \hat{\rho},\hat{H} ]=0$. Then, we perform splitting
\begin{eqnarray}
&&\mbox{\hspace{-5mm}}\hat{H} = \hat{H}_{+} - \hat{H}_{-}\, , \non \\
[2mm] &&\mbox{\hspace{-5mm}}\hat{H}_{+} =
\frac{1}{4}\hat{\rho}^{-1} \left( \hat{\rho} +
\hat{H}\right)^{2}\! \! , \; \;\; \hat{H}_{-} = \frac{1}{4}
{\hat{\rho}}^{-1}\left( \hat{\rho} - \hat{H}\right)^{2}\! \!,
\label{2.B.3}
\end{eqnarray}
where $\hat{H}_+$ and $\hat{H}_-$ are positive definite operators
satisfying
\bea [ \hat{H}_{+}, \hat{H}_{-}]\, = \,  [ \hat{\rho} , \hat{H} ]
= 0\, . \label{2.B.5}\eea
We may now employ the Dirac canonical quantization of constrained
systems and enforce a lower bound upon the Hamiltonian by imposing
the restriction
\begin{equation}
\hat{H}_{-}|\psi \ran = 0\, , \label{2.B.4}
\end{equation}
on the Hilbert space of {\em be-ables}. The resulting {\em
physical} state space, i.e. the space of {\em observables} has the
energy eigenvalues that are trivially positive owing to
\begin{eqnarray} \hat{H} |\psi \ran = \hat{H}_+ |\psi \ran = \hat{\rho} |\psi
\ran \, .
\end{eqnarray}
Concomitantly, in the Schr\"{o}dinger picture the equation of
motion
\begin{eqnarray}
\frac{d}{dt} |\psi_t \rangle = -i \hat{H}_+ |\psi_t \rangle\, ,
\end{eqnarray}
has only positive frequencies on physical states. Note that due to
condition (\ref{2.B.5}) 't~Hooft's constraint (\ref{2.B.4}) is a
first-class constraint. It is well known in the theory of
constrained dynamics~\cite{Sunder} that first-class conditions
generate gauge transformation and thus not only restrict the full
Hilbert space but also produce equivalence classes of states. It
should be noticed that above equivalence classes are generally
non-local, in the sense that two states belong to the same class
if they can be transformed into each other by gauge transformation
with the generator $\hat{H}_-$. If, in addition, the ensuing fiber
bundle structure is non-trivial one may encounter signatures of
this through the emergence of geometric phases.

't~Hooft proposed in Ref.~\cite{tHooft3} that in cases when the
dynamical equations (\ref{2.B.2a}) describe the
configuration-space chaotic dynamical system, the equivalent
classes could be related to its stable orbits (e.g., limit
cycles). The mechanism responsible for clustering of trajectories
to equivalence classes was identified by 't~Hooft as information
loss
--- after while one cannot retrace back the initial conditions of
a given trajectory, one can only say at what attractive trajectory
it will end up.
%
%
As the mechanism of equivalent classes is embodied in
Eq.(\ref{2.B.4}) we shall henceforth refer to it as {\em
information loss condition\/}. Applications of the the outlined
``canonical" scenario were given, e.g., in Refs.~\cite{BJV1}.

As Feynman's path integrals represent a legitimate alternative to
canonical quantization it is intriguing to formulate 't~Hooft's
procedure in the language of path integrals. Apart from the fact
that path integrals have a close proximity to classical physics,
they have also the additional advantage that they can incorporate
constraints
 in a straightforward manner. In this respect, the
Faddeev-Jackiw treatment of constrained systems is an interesting
option which we are going to explore in the following.



\section{Path integral quantization of 't~Hooft's system \label{SecIII}}

We now consider the path integral quantization \cite{PI} of the
class of systems described by Hamiltonians of the type
(\ref{2.B.1}). Because of the absence of a leading kinetic term
quadratic in the momenta $p_ a$, the system can be viewed as {\em
singular} and the ensuing quantization can be achieved through some
standard technique for quantization of constrained systems.

Particularly convenient is the technique   proposed by Faddeev and
Jackiw~\cite{F-J}. There one starts by observing that a Lagrangian
for 't~Hooft's equations of motion (\ref{2.B.2a}) and (\ref{2.B.2b})
can be simply taken as
\begin{eqnarray}
L({\bol{q}}, \dot{{\bol{q}}}, {\bol{p}}, \dot{\bol{p}}) ~= \
{\bol{p}}\cdot \dot{\bol{q}} - H({\bol{p}}, {\bol{q}})\, ,
\label{3.1}
\end{eqnarray}
with ${\bol{q}}$ and ${\bol{p}}$ being {\em Lagrangian variables\/}
(in contrast to  phase space variables). Note that $L$ does not
depend on $\dot{\bol{p}}$. It is easily verified that the
Euler-Lagrange equations for the Lagrangian (\ref{3.1}) indeed
coincide with the Hamiltonian equations (\ref{2.B.2a}) and
(\ref{2.B.2b}). Thus given 't~Hooft's Hamiltonian (\ref{2.B.1}) one
can always construct a first-order Lagrangian (\ref{3.1}) whose
configuration space coincides with the Hamiltonian phase space. By
defining $2N$ configuration-space coordinates as
\begin{eqnarray}
&&\xi^a ~= ~p_a, \;\;\; a = 1, \ldots, N\, ,\nonumber \\
&&\xi^a ~= ~q^a, \;\;\; a = N+1, \ldots, 2N\, ,
\end{eqnarray}
the Lagrangian (\ref{3.1}) can be cast into the more expedient
form, namely (summation convention understood)
\begin{eqnarray}
L({\bol{\xi}}, \dot{\bol{\xi}}) ~= ~\mbox{$\frac{1}{2}$} \xi^a
{\bol{\omega}}_{ab} \dot{\xi}^b - H({\bol{\xi}})\, . \label{3.26}
\end{eqnarray}
Here ${\bol{\omega}}$ is the $2N\times 2N$ symplectic matrix
\begin{eqnarray}
{\bol{\omega}}_{ab} ~= ~\lf(\ba{ll} 0&{\bol I}~\\-{\bol{I}}~&0
\ea\ri)_{ab}\, ,
\end{eqnarray}
which has an inverse ${\bol{\omega}}_{ab}^{-1} \equiv
{\bol{\omega}}^{ab}$. The equations of the motion read
\begin{eqnarray}
\dot{\xi}^a ~= ~{\bol{\omega}}^{ab} \frac{\partial
H({\bol{\xi}})}{\partial \xi^b}\, , \label{3.25}
\end{eqnarray}
indicating that there are no constraints on ${\bol{\xi}}$. Thus the
Faddeev-Jackiw procedure makes the system unconstrained, so that the
path integral quantization may proceed in the standard way. The time
evolution amplitude is simply \cite{PI}
%
\begin{eqnarray}
&&\mbox{\hspace{-6mm}}\langle {\bol{\xi}}_2,t_2| {\bol{\xi}}_1, t_1
\rangle \! =\! {\mathcal{N}} \int_{\boll{\xi}(t_1) =
\boll{\xi}_1}^{\boll{\xi}(t_2) = \boll{\xi}_2}\!
{\mathcal{D}}{\bol{\xi}} \ \exp \left[ \frac{i}{\hbar }\!
\int_{t_1}^{t_2}\! dt~L({\bol{\xi}}, \dot{\bol{\xi}})\right],
\nonumber
\\
&&\mbox{\hspace{-6mm}}\label{frad}
\end{eqnarray}
where ${\mathcal{N}}$ is some normalization factor, and the
measure can be rewritten as
\begin{equation}
{\mathcal{N}} \int_{\boll{\xi}(t_1) =
\boll{\xi}_1}^{\boll{\xi}(t_2) = \boll{\xi}_2}
{\mathcal{D}}{\bol{\xi}} = {{\mathcal{N}}} \int_{{\boll{q}}(t_1) =
{\boll{q}}_1}^{{\boll{q}}(t_2) = {\boll{q}}_2}
{\mathcal{D}}{\bol{q}} {\mathcal{D}}{{\bol{p}}} \, .
\label{15}\end{equation}
Since the Lagrangian
(\ref{3.1})
 is linear in ${\bol{p}}$,
 we may integrate these variables out and obtain
\begin{eqnarray}
&&\!\!\!\!\!\!\!\!\!\!\!\!\!\!\!\!\!\langle {\bol{q}}_2,t_2|
{\bol{q}}_1, t_1 \rangle = {{\mathcal{N}}} \int_{{\boll{q}}(t_1) =
{\boll{q}}_1}^{{\boll{q}}(t_2) = {\boll{q}}_2}
{\mathcal{D}}{\bol{q}} ~\prod_a \delta[ \dot{q}^a-f^a({\bol{q}})]\,
, \label{eg.1.2}
\end{eqnarray}
 where
$ \delta[{\bol{f}} ]\equiv \prod_t  \delta ({\bol{f}}(t))$ is the
functional version of Dirac's   $ \delta $-function. Hence, the
system described by the Hamiltonian (\ref{2.B.1}) retains its
deterministic character even after quantization. The paths are
squeezed onto the classical trajectories determined by the
differential equations
 $\dot{\bol{q}} = {\bol{f}}({\bol{q}})$. The time evolution
amplitude (\ref{eg.1.2}) contains a sum  over only the classical
trajectories --- there are no quantum fluctuations driving the
system away from the classical paths, which is precisely what should
be expected from a deterministic dynamics.

The amplitude (\ref{eg.1.2}) can be brought into more intuitive form
by  utilizing the identity
\begin{eqnarray}
\delta\left[ {\bol{f}}({{\bol{q}}})  - \dot{{\bol{q}}} \right] ~=
~\delta[ {\bol{q}} - {\bol{q}}_{\rm cl}]~(\det {{M}})^{-1}\, ,
\end{eqnarray}
where ${M}$ is a functional matrix formed by the second functional
derivatives of the action ${\calS}[{\bol{\xi}}]\equiv \int
dt\,L({\bol{\xi}}, \dot{\bol{\xi}}) $\,:
\begin{eqnarray}
{{M}}_{ab}(t,t') ~= ~\left. \frac{\delta^2 {\calS}}{\delta
\xi^a(t)~\delta {\xi}^b(t')}~\right|_{{\bol{q}} = {\bol{q}}_{\rm
cl}} \, . \label{4.01}
\end{eqnarray}
The Morse index theorem  ensures that for sufficiently short
time intervals $t_2-t_1$ (before the system reaches its first
focal point), the classical solution with the initial condition
${{\bol{q}}}(t_1) = {\bol{q}}_1$ is unique.
In such a case, Eq.~(\ref{eg.1.2}) can be brought to the form
\begin{eqnarray}
\langle {\bol{q}}_2,t_2| {\bol{q}}_1, t_1 \rangle &=&
{\tilde{\mathcal{N}}} \int_{{\boll{q}}(t_1) =
{\boll{q}}_1}^{{\boll{q}}(t_2) = {\boll{q}}_2}
{\mathcal{D}}{\bol{q}} ~\delta\left[{\bol{q}} - {\bol{q}}_{\rm
cl} \right]\, , \label{4.2}
\end{eqnarray}
with ${\tilde{\mathcal{N}}}\equiv {{\mathcal{N}}}/(\det {M})$.
Remarkably,
the Faddeev-Jackiw treatment bypasses completely
the discussion of
constraints, in contrast
with the conventional Dirac-Bergmann method~\cite{Dir2,Sunder}
where $2N$ (spurious) second-class primary constraints must be
introduced to deal with 't~Hooft's system, as done in \cite{BJK}.

\section{Emergent SUSY --- signature of
classicality}

We now turn to an interesting implication of the result (\ref{4.2}).
If we had started in Eq.(\ref{eg.1.2}) with an external current
\begin{eqnarray}
\tilde L({\bol{\xi}}, \dot{\bol{\xi}}) =
 L({\bol{\xi}}, \dot{\bol{\xi}}) + i\hbar {\bol{J}}\cdot {\bol{q}}\,
,
\end{eqnarray}
integrated again over ${\bol{p}}$, and took the trace over
${\bol{q}}$, we would end up with a generating functional
\begin{eqnarray}
{\mathcal{Z} }_{\rm CM}[\bol{J}] = \tilde{{\mathcal{N}}} \int
{\mathcal{D}}{\bol{q}} ~\delta[{\bol{q}}- {\bol{q}}_{\rm cl}]
 \exp\left[\int_{t_1}^{t_2} dt \ {\bol{J}}\cdot{\bol{q}}\right]\, .
\label{4.3}
\end{eqnarray}
This coincides with the path integral formulation of classical
mechanics postulated by Gozzi {\em et al}.~\cite{GozziI,GozziII}.
The same representation can be derived
from the classical limit of a
closed-time path integral for the
transition probabilities of a quantum particle in a
heat bath~\cite{PI,BJK},
The path integral
  (\ref{4.3})
has an interesting mathematical
structure. We may rewrite it as
\begin{eqnarray}
{\mathcal{Z} }_{\rm CM}[\bol{J}] ~&=& ~\tilde{{\mathcal{N}}} \int
{\mathcal{D}}{\bol{q}} ~\delta\left[ \frac{\delta {\calS}}{\delta
{\bol{q}}} \right] ~\det \left| \frac{\delta^2 {\calS} }{\delta q_a
(t) ~q_b(t')} \right|
\nonumber \\
&\times& ~\exp\left[\int_{t_1}^{t_2} dt \
{\bol{J}}\cdot{\bol{q}}\right]\, . \label{4.4}
\end{eqnarray}
By representing the delta functional in the usual way as a
functional Fourier integral
\begin{eqnarray*}
\delta\left[ \frac{\delta {\calS}}{\delta {\bol{q}}} \right] = \int
{\mathcal{D}} {\bol{\lambda}} ~\exp\left( i \int_{t_1}^{t_2} dt
~{\bol{\lambda}}(t) \frac{\delta {\calS}}{\delta {\bol{q}}(t)}
\right)\, ,
\end{eqnarray*}
and the functional determinant as a functional integral over two
real time-dependent Grassmannian {\em ghost variables\/} $c_a(t)$
and $\overlinen{c}_a(t)$,
\begin{eqnarray*}
&&\det \left| \frac{\delta^2 \calS }{\delta q^a (t) ~\delta q^b(t')}
\right|\nonumber \\
&&= \int {\mathcal{D}}{\mathbf{c}} {\mathcal{D}}
\overlinen{{\bol{c}}} ~\exp{\left[ \int_{t_1}^{t_2}dt dt' \
\bar{c}_a(t) \frac{\delta^2 {\calS} }{\delta{q^a} (t) \
\delta{q^b}(t')} ~{c_b}(t')\right]}\, ,
\end{eqnarray*}
we obtain
\begin{eqnarray}
&&\mbox{\hspace{-10mm}}{\mathcal{Z} }_{\rm CM} [\bol{J}] =  \int\!
{\mathcal{D}}{\bol{q}}{\mathcal{D}}{\bol{\lambda}}{\mathcal{D}}{\bol{c}}
{\mathcal{D}}\bar{{\bol{c}}}  \exp\left[ i {\mathcal{S}} +
\int_{t_1}^{t_2} \! dt ~{{\bol{J}}}\cdot{{\bol{q}}} \right]\! ,
\label{4.5a}
\end{eqnarray}
with the new action
\begin{eqnarray}
&&\mbox{\hspace{-7mm}}{\mathcal{S}}[{\bol{q}}, \bar{{\bol{c}}},
{\bol{c}}, {\bol{\lambda}}] \equiv ~\int_{t_1}^{t_2} dt \
{\bol{\lambda}}(t)\frac{\delta {\calS}}{\delta
{\bol{q}}(t)}\nonumber \\ &&\mbox{\hspace{-2mm}}-
i\int_{t_1}^{t_2}dt \int_{t_1}^{t_2} dt' ~\bar{c}_a(t)
\frac{\delta^2 {\calS} }{\delta{q^a} (t) ~\delta{q^b}(t')} \
{c_b}(t')\, . \label{4.5}
\end{eqnarray}
Since ${\mathcal{Z} }_{\rm CM}[\bol{J}]$ can be derived from
the classical limit of a closed-time path integral
for the transition  probability,
it comes to no
surprise that ${\mathcal{S}}$ exhibits BRST (and anti-BRST)
symmetry.
It is simple to check~\cite{BJK} that $\mathcal{S}$ does
not change under  the symmetry transformations
\begin{eqnarray}
&&\delta_{{\rm BRST\,}}  {\bol{q}} = \bar{\varepsilon} {\bol{c}}\,
,
 \;\;\; \delta_{{\rm BRST\,}}  \bar{{\bol{c}}} =
-i\bar{\varepsilon} {\bol{\lambda}}\, , \;\;\; \delta_{{\rm
BRST\,}} {\bol{c}} = 0\, , \nonumber \\&&\delta_{{\rm BRST\,}}
{\bol{\lambda}} = 0\, , \label{4.6}
\end{eqnarray}
where $\bar{\varepsilon}$ is a Grassmann-valued parameter (the
corresponding anti-BRST transformations are related to
(\ref{4.6}) by charge conjugation).
%
%
\noi
As noted in~\cite{GozziII}, the ghost fields $\bar{{\bol{c}}}$ and
${\bol{c}}$ are mandatory at the classical level as their r\^{o}le
is to cut off the fluctuations {\em perpendicular\/} to the
classical trajectories. On the formal side, $\bar{{\bol{c}}}$ and
${\bol{c}}$ may be identified with Jacobi
fields~\cite{GozziII,DeWitt}. The corresponding BRST charges
are related to Poincar\'{e}-Cartan integral
invariants~\cite{GozziIII}.

By analogy with the stochastic quantization the path integral
(\ref{4.5a}) can be rewritten in a compact form with the help of a
superfield~\cite{GozziI,Zinn-JustinII,PI}
\begin{eqnarray}
&&\mbox{\hspace{-10mm}}\Phi_a(t, \theta, \bar{\theta}) = ~q_a(t) +
i\theta c_a(t) -i\bar{\theta} \bar{c}_a(t) + i
\bar{\theta}\theta \lambda_a(t)\, , \label{3.23}
\end{eqnarray}
in which $\theta$ and $\overlinen{\theta}$ are anticommuting
coordinates extending the configuration space of ${\bol{q}}$
variables to a superspace. The latter is nothing but the degenerate
case of supersymmetric field theory in $d=1$ in the superspace
formalism of Salam and Strathdee~\cite{SS1}. In terms of superspace
variables we see that
\begin{eqnarray}
&&\!\!\!\!\!\!\!\!\!\!\!\int d\bar{\theta} d\theta ~{\calS}[{\bol{\Phi}}]\\
&&\!\!\!\!\!\!\!\!\!\!\!\!\!\!\!\!\!\mbox{\hspace{5mm}}= \int dt d\bar{\theta} d\theta \
L({\bol{q}}(t) + i\theta {\bol{c}}(t) - i \bar{\theta}
\bar{\bol{c}}(t) + i \bar{\theta}\theta \bol{\lambda}(t)
) =-i {\mathcal{S}}\nonumber
\end{eqnarray}
To obtain the last line we Taylor expanded $L$ and used the
standard integration rules for Grassmann variables. Together with
the identity ${\mathcal{D}} {\bol{\Phi}} =
{\mathcal{D}}{\bol{q}}{\mathcal{D}}{\bol{c}}{\mathcal{D}}\bar{{\bol{c}}}
{\mathcal{D}}{\bol{\lambda}}$ we may therefore express the
classical partition functions (\ref{4.3}) and (\ref{4.4}) as a
supersymmetric path integral with fully fluctuating paths in
superspace
\begin{eqnarray*}
{\mathcal{Z} }_{\rm CM}[\bol{J}]  &=&  \int\! {\mathcal{D}}
{\bol{\Phi}} \
\exp\left\{- \int d\theta d\bar{\theta} \
{\calS}[{\bol{\Phi}}](\theta, \bar{\theta}) \right\} \nonumber \\
&& \times  ~\exp\left\{\int dt d\theta d\bar{\theta} \
{\bol{\Gamma}}(t, \theta, \bar{\theta}){\bol{\Phi}}(t, \theta,
\bar{\theta})\right\}\, . \label{3.24}
\end{eqnarray*}
Here we have introduced the supercurrent ${\bol{\Gamma}}(t,
\theta, \bar{\theta}) = \bar{\theta} \theta {\bol{J}}(t)$.

Let us finally add that under rather general assumptions it is
possible to prove~\cite{BJK} that 't~Hooft's deterministic systems
are the {\em only} systems with the peculiar property that their
full quantum properties are classical in
the Gozzi {\em et al.} sense.
Among others, the latter also indicates that the
Koopman-von~Neumann operator formulation of classical
mechanics~\cite{KN1} when applied to 't~Hooft systems must agree
with their canonically quantized counterparts.

\section{Inclusion of information loss}

As observed in Section~II, the Hamiltonian (\ref{2.B.1}) is not
bounded from below, and this is clearly true for any function
$f^a({\bol{q}})$. Hence, no deterministic system with dynamical
equations $\dot{q}^a = f^a({\bol{q}})$ can describe a stable {\em
quantum world\/}. To deal with this situation we now employ
't\,Hooft's procedure of Section~II. We assume that the system
(\ref{2.B.1}) has $n$ conserved irreducible charges $C^i$, i.e.,
\begin{eqnarray}
\{ C^i, H \} = 0\, , \;\;\;\; i = 1, \ldots, n\, . \label{5.1}
\end{eqnarray}
Then, we enforce a lower bound upon $H$, by imposing the condition
that $H_-$ is zero on the physically accessible part of a phase
space.

The splitting  of $H$ into $H_-$ and $H_+$ is conserved in time
provided that $\{ H_-, H \} = \{ H_+, H \} = 0$, which is
ensured if
 $\{ H_+, H_- \} = 0$. Since the charges $C^i$ in
(\ref{5.1}) form an irreducible set,
 the Hamiltonians
 $H_+$
and $H_-$ must be functions of the charges and $H$ itself.
 There is a certain amount of
flexibility in finding $H_-$ and $H_+$. For convenience take the
following choice:
\begin{eqnarray}
H_+ ~= ~\frac{(H + a_i C^i)^2}{4  a_i C^i} \, , \; \; H_- ~= \
\frac{(H - a_i C^i)^2}{4  a_i C^i} \, , \label{FCH}
\end{eqnarray}
where $a_i(t)$ are ${\bol{q}}$ and ${\bol{p}}$ independent. The
lower bound is reached
 by choosing $a_i(t) C^i$ to be non-negative.
We shall select a combination of $C^i$ which is
${\bol{p}}$-independent [this condition may not necessarily be
achievable for general $f^a({\bol{q}})$].

In the Dirac-Bergmann quantization  approach used in our previous
paper \cite{BJK}, the information loss condition (\ref{2.B.4}) was a
first-class primary constraint. In the Dirac-Bergmann analysis, this
signals the presence of a gauge freedom --- the associated Lagrange
multipliers cannot be determined from dynamical equations
alone~\cite{Dir2}. The time evolution of observable quantities,
however, should not be affected by the arbitrariness of Lagrange
multipliers. To remove this superfluous freedom one must choose a
gauge. For details of this more complicated procedure see
\cite{BJK}.

In the Faddeev-Jackiw approach, Dirac's elaborate classification
of constraints to first or second class, primary or secondary is
avoided. It is therefore worthwhile to rephrase the entire
development in Ref.~\cite{BJK} once more in this approach. The
information loss condition may now be introduced by simply adding
to the Lagrangian (\ref{3.26}) a term enforcing
\begin{eqnarray}
H_-({\bol{\xi}}) ~= ~0\, , \label{5.3}
\end{eqnarray}
by means of a Lagrange multiplier. More in general we can take
instead of $H_-$ any function $\phi({\bol{\xi}})$, such that
$\phi({\bol{\xi}}) = 0$ implies $H_-({\bol{\xi}}) = 0$. In this
way we obtain
\begin{eqnarray}
L({\bol{\xi}}, \dot{\bol{\xi}})  =  \mbox{$\frac{1}{2}$} \xi^a
{\bol{\omega}}_{ab} \dot{\xi}^b - H({\bol{\xi}}) - \eta \ \!
\phi({\bol{\xi}})\, , \label{5.2}
\end{eqnarray}
In Faddeev-Jackiw method one directly applies the constraint and
thus eliminates one of $\xi^a$, say $\xi^1$, in terms of the
remaining coordinates. This reduces the dynamical variables to
$2N-1$. Apart from an irrelevant total derivative, the canonical
term $ \xi^a {\bol{\omega}}_{ab} \dot{\xi}^b$ changes to $ \xi^i
{\bol{f}}_{ij}(\hat{{\bol{\xi}}})\dot{\xi}^j$, with
\begin{equation}
 {\bol{f}}_{ij}(\hat{{\bol{\xi}}})= {\bol{\omega}}_{ij}- \left[ {\bol{\omega}}
_{1i}\frac{\partial \xi^1}{\partial \xi^j}-(i \leftrightarrow
j)\right]. \label{32}\end{equation}
Here $i,j = 2, \ldots, 2N$, and $\hat{{\bol{\xi}}} = \{\xi^2,
\ldots, \xi^{2N} \}$. Eliminating $\xi^1$ also in the Hamiltonian
$H$ we obtain the reduced Hamiltonian $H_R(\hat{\bol{\xi}})$,   so
that we are left with the reduced Lagrangian
\begin{eqnarray}
L_R(\hat{{\bol{\xi}}}, \dot{\hat{\bol{\xi}}})  ~= \
\mbox{$\frac{1}{2}$}\xi^i {\bol{f}}_{ij}(\hat{{\bol{\xi}}})
\dot{\xi}^j - H_R(\hat{{\bol{\xi}}})\, . \label{5.34}
\end{eqnarray}
At this point  one must worry about the notorious
operator-ordering problem, not knowing in which temporal order
$\hat{{\bol{\xi}}}$ and $\dot{\hat{\bol{\xi}}}$ must be taken in
the kinetic term. A path integral in which the kinetic term is
coordinate-dependent can in general only be defined
perturbatively, in which all anharmonic terms are treated as
interactions. The partition function is expanded in powers of
expectation values of products of these interactions which, in
turn, are expanded into integrals over all Wick contractions, the
Feynman integrals. Each contraction represents a Green function.
For the Lagrangian of the form (\ref{5.34}), the contractions of
two ${{{\xi}^i}}$'s contain  a Heaviside step function, those of
one ${{{\xi}^i}}$ and one $\dot{{{\xi}^i}}$ contain a Dirac $
\delta $-function, and those of two $\dot{{{\xi}^i}}$'s contain a
function $\dot  \delta (t-t')$. Thus, the Feynman integrals run
over products of distributions and are mathematically undefined.
Fortunately, a unique definition has recently been found. It is
enforced by the necessary physical requirement that path integrals
must be invariant under coordinate transformations \cite{KC}.

The Lagrangian is processed further with the help of Darboux's
theorem~\cite{Darboux}. This allows us to perform a non-canonical
transformation $\xi^i \mapsto (\zeta^s, z^r)$ which brings $L_R$
to the canonical form
\begin{eqnarray}
L_R({\bol{\zeta}, \dot{\bol{\zeta}}}, {\bol{z}}) ~= \
\mbox{$\frac{1}{2}$}\zeta^s {\bol{\omega}}_{st}\dot{\zeta}^t -
H_R'({\bol{\zeta}}, {\bol{z}})\, , \label{5.35}
\end{eqnarray}
where ${\bol{\omega}}_{st}$ is the canonical symplectic matrix in
the reduced $s$-dimensional space. Darboux's theorem ensures that
such a transformation exists at least locally. The variables $z^r$
are related to zero modes of the matrix
${\bol{f}}_{ij}(\hat{{\bol{\xi}}})$ which makes it non-invertible.
Each zero mode corresponds to a constraint of the system. In
Dirac's language these would correspond to the secondary
constraints. Since there is no $\dot z^r$ in the Lagrangian, the
variables $z^r$ do not play any dynamical r\^{o}le and can be
eliminated using the equations of motion
\begin{eqnarray}
\frac{\partial H_R'({\bol{\zeta}}, {\bol{z}})}{\partial z^r} ~= \
0\, . \label{5.36}
\end{eqnarray}
In general, $H_R'({\bol{\zeta}}, {\bol{z}})$ is a nonlinear function
of $z^{r_1}$. One now solves as many $z^{r_1}$ as possible in terms
of remaining $z$'s, which we label by $z^{r_2}$, i.e.,
\begin{eqnarray}
z^{r_1} ~= ~\varphi^{r_1}({\bol{\zeta}}, z^{r_2})\, . \label{5.4}
\end{eqnarray}
If $H_R'({\bol{\zeta}}, {\bol{z}})$ happens to be linear in
$z^{r_2}$, we obtain the constraints
\begin{eqnarray}
\varphi_{r_2}({\bol{\zeta}}) ~= ~0\, . \label{5.5}
\end{eqnarray}
Inserting the constraints (\ref{5.4}) into (\ref{5.35}) we obtain
\begin{eqnarray}
L_R({\bol{\zeta}}, \dot{\bol{\zeta}}, {\bol{z}}) ~= \
\mbox{$\frac{1}{2}$}\zeta^s {\bol{\omega}}_{st}\dot{\zeta}^t -
H_R''({\bol{\zeta}}) - z^{r_2} \varphi_{r_2}({\bol{\zeta}})\, ,
\end{eqnarray}
with $z^{r_2}$ playing the r\^{o}le of Lagrange multipliers. We now
repeat the elimination procedure until there are no more
$z$-variables. The surviving variables represent the true physical
degrees of freedom. In the Dirac-Bergmann approach, these would span
the {\em reduced} phase space $\Gamma^*$.

Let us follow the procedure in more detail if there is just one
variable $z$ in (\ref{5.36}) and only equation (\ref{5.4}) holds.
As in Ref.~\cite{BJK}, we can pass to the new set of canonical
variables ${\bol{\xi}} \mapsto ({\bol{\zeta}}, z, p_z)$ with $p_z
= \phi$. Let us define the function
\begin{eqnarray}
\chi({\bol{\zeta}}, z) & \equiv & \frac{\partial H_R'({\bol{\zeta}},
z)}{\partial z} ~= ~ \frac{\partial H_+(\xi^1(\hat{\bol{\xi}}),
\hat{\bol{\xi}})}{\partial z} \nonumber \\
&=& \left. \{ H_+, \phi \} \right|_{p_z = 0} ~= ~ 0\, .
\label{5.37}
\end{eqnarray}
Its derivative is given by the Poisson bracket
\begin{eqnarray} \!\!
\frac{\partial \chi({\bol{\zeta}}, z)}{\partial z} \ = \
\{\chi({\bol{\zeta}}, z), p_z\}\ =\  \{\chi, \phi \}  \ \neq \ 0 \,
.. \label{5.38}
\end{eqnarray}
Because (\ref{5.38}) is different from zero on account of
(\ref{5.4}) we can identify the function $\chi({\bol{\zeta}}, z)$
with the implicit gauge fixing condition of the Faddeev-Jackiw
analysis.

Let us now see how we can include the constraints (\ref{5.3}) and
(\ref{5.37}) into 
the path integral (\ref{4.3}) for ${\mathcal{Z}}_{\rm
CM}[\bol{J}]$. This cannot simply be done by inserting
$\delta$-functionals $\delta[\phi]$ and $\delta[\chi]$ into the
integrand, since $\phi$ and $\chi$ may not be independent.
Allowing for this, the path integral reads (see~Ref.\cite{BJK})
\begin{eqnarray}
{\mathcal{Z}}_{\rm CM}[\bol{J}] \!\! &=& \int
{\mathcal{D}}{\bol{\xi}}\ \delta[\phi] \delta[\chi] |\det|\!| \{
\phi,\chi\} |\!| |
\nonumber\\
&\times& ~\exp\left[i\!\!\int^{t_f}_{t_i}\!\!dt ~L({\bol{\xi}},
\dot{\bol{\xi}}) + \int^{t_f}_{t_i}\!\!dt ~{\bol{J}}{\bol{\xi}}
\right]. \label{V41}
\end{eqnarray}
Assuming that $\xi^1$ can be eliminated globally from (\ref{5.2}),
we obtain
\begin{eqnarray}
{\mathcal{Z}}_{\rm CM}[\bol{J}] \!\! \!&=&\!\!\!\displaystyle \int
{\mathcal{D}}\hat{\displaystyle \bol{\xi}}\ \displaystyle
\delta[\chi]|\det|\!| \{ \phi,\chi\} |\!|
|\left|\det\left|\!\left|\frac{\delta \phi}{\delta
\xi^1}\right|\!\right|\right|^{-1}_{\xi^1 =
\xi^1(\hat{{\boll{\xi}}})}
\nonumber\\
\!\! &\times & \!\!\!  \exp\left[i \!\!\int^{t_f}_{t_i}\!\!dt \
L_R(\hat{{\bol{\xi}}}, \dot{\hat{\bol{\xi}}}) +
\int^{t_f}_{t_i}\!\!dt
~{\bol{J}}{\bol{g}}(\hat{{\bol{\xi}}})\right]\!.
\end{eqnarray}
After the Darboux transformation, this becomes
\begin{eqnarray}\!\!\!\!\!\!\!\!\!\!\!
{\mathcal{Z}}_{\rm CM}[\bol{J}] \!\! &=& \!\!\! \int
{\mathcal{D}}{\bol{\zeta}} {\mathcal{D}}z~\delta[z -
\varphi({\bol{\zeta}})]\nonumber \\
\!\! &\times & \!\!\!  \exp\left[i\!\!\int^{t_f}_{t_i}\!\!dt \
L_R({{\bol{\zeta}}}, \dot{{\bol{\zeta}}},z)\! + \!
\int^{t_f}_{t_i}\!\!dt
~{\bol{J}}{\bol{g}}({{\bol{\zeta}}},z)\right]\!,
\end{eqnarray}
where we have used the functional relation
\begin{eqnarray}
\delta[\chi] |\det|\!| \{ \phi, \chi \} |\!| | &=& \delta\!\left[
\frac{\delta H_+}{\delta z}\right]\left|\det\left|\!\left|
\frac{\delta^2H_+}{\delta z(t) \delta z(t')} \right|\!\right|
\right|\nonumber \\
&=& \delta[z - \varphi({\bol{\zeta}})]\, ,
\end{eqnarray}
together with Jacobi-Liouville equality
\begin{eqnarray}
&&\frac{\partial(\xi^2, \ldots, \xi^{2N})}{\partial(\zeta^1,
\ldots, \zeta^{2N-2},z)}\nonumber \nonumber \\
&&\mbox{\hspace{3mm}}  = \frac{\partial(\xi^2, \ldots, \xi^{2N},
p_z)}{\partial(\zeta^1, \ldots, \zeta^{2N-2},z,
p_z)}\frac{\partial(\zeta^1, \ldots, \zeta^{2N-2}, z,
p_z)}{\partial(\xi^2, \ldots, \xi^{2N}, \xi^1)}\nonumber \\
&&\mbox{\hspace{3mm}}  = \left(\frac{\partial p_z}{\partial
\xi^1}\right)_{\!\hat{\boll{\xi}}} = \left(\frac{\partial
\phi}{\partial \xi^1}\right)_{\!\xi^1 = \xi^1(\hat{\boll{\xi}})}\,
..
\end{eqnarray}
With the notation $H_+^*({\bol{\zeta}}) = H_+({\bol{\zeta}}, z =
\varphi({\bol{\zeta}}), p_z = 0 )$, this can be rewritten as
\begin{eqnarray} \!\!\!\!\!\!\!\!\!\!\!\!
{\mathcal{Z}}_{\rm CM}[\bol{J}] \!\! &=& \!\!\! \int
{\mathcal{D}}{\bol{\zeta}} \exp\left[i \!\!\int^{t_f}_{t_i}\!\!dt
~\zeta^t {\bol{\omega}}_{ts} \dot{\zeta}^s \right] \nonumber \\
\!\! &\times & \!\!\! \exp\left[-i\!\!\int^{t_f}_{t_i}\!\!dt \
H_+^*({\bol{\zeta}})  + \! \int^{t_f}_{t_i}\!\!dt \
{\bol{J}}{\bol{g}}^*({{\bol{\zeta}}})\right]. \label{5.41}
\end{eqnarray}
At this point we note that the result (\ref{5.41}) is equivalent
to the result derived in Ref.~{\cite{BJK}}. In fact, when $\chi$
in~{\cite{BJK}} coincides with the the form (\ref{5.37}) and we
set ${\bol{\zeta}} = (\bar{\bol{Q}}, \bar{\bol{P}})$, $z = Q_1$,
and $p_z = P_1$, then ${\mathcal{Z}}_{\rm CM}[\bol{J}]$ from
Ref.~{\cite{BJK}} reduces exactly  to the form (\ref{5.41}). In
general cases, however, the gauge fixing condition of the
Dirac-Bergmann procedure can be chosen in a different way with respect to
the natural choice implicit in the Faddeev-Jackiw analysis. In
such a situation the resulting reduced Lagrangians do not coincide
but are connected via a canonical transformation.

Important simplification happens when $H'_R$ is independent of $z$
(e.g., when $\phi = H_-$). In Dirac-Bergmann's language this
indicates that there is no secondary constraint. In such a case
the gauge fixing can be enforced by choosing $\chi = z$ (see
L.~Faddeev in Ref.~\cite{Fad}), and the procedure outlined in
steps (\ref{V41})- (\ref{5.41}) is streamlined by the fact that
$|\det|\!| \{ \phi,\chi\} |\!| | =1$. The corresponding coordinate
basis $\{ {\bol{\zeta}}, \chi, \phi\}$ is known as the
Shouten-Eisenhart basis~\cite{Sunder}.

\section{Examples of emergent
quantum systems}
\subsection{Free particle}

We conclude our presentation by exhibiting how our quantization
method works for a simple system described by 't~Hooft's Hamiltonian
\begin{eqnarray}
H({\bol{q}}, {\bol{p}}) \ = \ xp_y - yp_x\, . \label{6.1}
\end{eqnarray}
Formally, this represents the $z$ component of the angular
momentum, whose spectrum is unbounded from below. Alternatively,
one can regard (\ref{6.1}) as describing the mathematical
pendulum. This is because the corresponding dynamical equation
(\ref{2.B.2a}) for ${\bol{q}}$ is a plane pendulum equation with
the pendulum constant $l/\mbox{{\textsl{g}}} =1$. The Lagrangian
(\ref{3.1}) then reads
\begin{eqnarray}
L({\bol{q}}, \dot{\bol{q}}, {\bol{p}}, \dot{\bol{p}}) ~= ~p_x
\dot{x} + p_y \dot{y} - xp_y + yp_x\, .
\end{eqnarray}
Here, indeed, the $L$ is $\dot{\bol{p}}$-independent, as discussed
in Section~III. It is well-known~\cite{Lutzky} that the system
(\ref{6.1}) has two  independent constants of motion - the Casimir
functions:
\begin{eqnarray}
C_1 ~= ~x^2 + y^2\, , \;\;\; C_2 ~= ~x p_x + y p_y\, .
\end{eqnarray}
Only $C_1$ is $\bol{p}$-independent, so that 't~Hooft's constraint
$\rho({\bol{q}})$ acquires the form:  $\rho({\bol{q}}) = a_1
C_1({\bol{q}})$, with the constant $a_1$ to be determined later.

The Faddeev-Jackiw procedure is based on the reduced Lagrangian
\begin{eqnarray}
&&\mbox{\hspace{-4mm}}L_R(\hat{{\bol{\xi}}},
\dot{\hat{\bol{\xi}}}) \  =  \ \dot{y}p_y + \frac{\dot{x}}{y} (p_y
x - a_1 (x^2 + y^2)) -
a_1(x^2 + y^2)\nonumber \\
&&~ \nonumber \\ &&\mbox{\hspace{10mm}} = \, \sqrt{(x^2 +
y^2)}~\frac{d}{dt}\! \left[-2 a_1  \sqrt{(x^2 + y^2)} \
\mbox{arct}\!\!\left(
\frac{x}{y}\right)\right.\nonumber \\
&& \mbox{\hspace{10mm}}\left. - \, \frac{x p_x + y p_y}{\sqrt{(x^2
+ y^2)}} \right] - \,  a_1(x^2 + y^2)\, . \label{6.3}
\end{eqnarray}
We can diagonalize the symplectic structure by means of the
Darboux transformation
\begin{eqnarray}
&&\mbox{\hspace{-10mm}}p_{\zeta} \ = \ \sqrt{x^2 + y^2} \, , \nonumber \\
&&\mbox{\hspace{-10mm}}\zeta \ =  \ - 2 a_1 \sqrt{(x^2 + y^2)} \
\mbox{arct}\!\!\left( \frac{x}{y}\right) - \, \frac{x p_x + y
p_y}{\sqrt{(x^2 + y^2)}} \; . \label{VI51}
\end{eqnarray}
Thus, up to a total derivative, the reduced Lagrangian (\ref{6.3})
goes over into
\begin{eqnarray}
L_R({\bol{\zeta}}, \dot{\bol{\zeta}}, z)  \ = \
\mbox{$\frac{1}{2}$}\zeta^s {\bol{\omega}_{st}}{\dot{\zeta}}^t \ -
\ a_1 (p_{\zeta})^2 \, , \label{6.4}
\end{eqnarray}
with the symplectic notation $\zeta \equiv \zeta^1$ and $p_{\zeta}
\equiv\zeta^2$. The reduced Hamiltonian is $z$-independent and
thus $\chi = z$. Note that (\ref{VI51}) together with
\begin{eqnarray}
z   =   - \mbox{arct}\!\!\left( \frac{x}{y}\right), \;\;\;\;
\mbox{and} \;\;\;\; p_z^2  =  \phi^2  =  4a_1 p_{\zeta}^2 \ \! H_-
\, , \label{VI52}
\end{eqnarray}
constitute the canonical transformation ${\bol{\xi}} \mapsto
({\bol{\zeta}}, z, p_z)$.

Due to a non-linear nature of the canonical transformation
(\ref{VI51}) and (\ref{VI52}) we must check up the path integral
measure for a potential anomaly. In Appendix~A, we show that
although the anomaly is indeed generated, it gets cancelled due to
the presence of the constraining $\delta$--functionals in the
measure. In other words, the Liouville anomaly is not present in the
reduced phase space. In addition, because (\ref{6.4}) is cyclic in
$\zeta$, it can be argued~\cite{SW} that no new (non-classical)
corrections to the Hamiltonian are generated in the action after the
above canonical transformation is performed.

Let us now set $a_1 = 1/2m\hbar$. After changing in the path
integral the variable $\zeta(t)$ to $\zeta(t)/\hbar$ we obtain the
path integral measure of quantum systems:
\begin{eqnarray}
{\mathcal{D}}{\bol{\zeta}} ~\approx ~\prod_i \left[\frac{ d
{\zeta}(t_i) d {p_{\zeta}}(t_i)}{2\pi \hbar}\right]\, .
\end{eqnarray}
In addition, the prefactor $1/\hbar$ in the exponent emerges
correctly. Thus, the classical partition function of Gozzi {\em et
al.} turns into the quantum partition function for a free particle
of mass $m$. As the constant $a_1$ represents the choice of units
(or scale factor) for $C_1$ we see that the quantum scale $\hbar$ is
implemented into the partition function via the choice of the
information loss  condition.

A free particle can emerge also from another class of 't~Hooft's
systems. Such systems can be obtained by modifying slightly the
previous discussion and considering instead the Hamiltonian
\begin{eqnarray}
H ~= ~xp_y - yp_x + \lambda(x^2 + y^2)\, ,
\end{eqnarray}
where $\lambda$ is a constant. 't~Hooft's information loss condition
and $\rho({\bol{q}})$ remain clearly  the same as in the previous
case. The reduced Lagrangian then reads
\begin{eqnarray}
&&\mbox{\hspace{-5mm}}L_R(\hat{{\bol{\xi}}},
\dot{\hat{\bol{\xi}}}) = \, \sqrt{(x^2 + y^2)}~\frac{d}{dt}\!
\left[- 2 a_1  \sqrt{(x^2 + y^2)} \ \mbox{arct}\!\!\left(
\frac{x}{y}\right)\right.\nonumber \\
&& \mbox{\hspace{8mm}}\left. - \, \frac{x p_x + y p_y}{\sqrt{(x^2
+ y^2)}} \right] - \,  a_1^*(x^2 + y^2)\, .
\end{eqnarray}
with $a_1^* = a_1 + \lambda$. Identical reasonings as in the
preceding situation lead again to a proper quantum-mechanical
partition function for a free particle.

\subsection{Harmonic oscillator}

In a previous paper that utilized the Dirac-Bergmann
treatment~\cite{BJK}, it was shown that the system (\ref{6.1}) can
be also used to obtain the quantized linear harmonic oscillator.
This is because there is a certain ambiguity in imposing 't~Hooft's
condition. This will be illustrated with $\phi = xp_y - yp_x -
a_1(x^2 + y^2)$ used in Eq.(\ref{VI52}). The constraint $\phi = 0$
can be equivalently written as
\begin{eqnarray}
\phi \ = \ {\bol{x}}\wedge {\bol{A}} \ = \ 0\, ,
\end{eqnarray}
with ${\bol{x}} = (x,y)$ and ${\bol{A}} = (p_x + a_1y, p_y -
a_1x)$. The solution of $\phi = 0$ is formally given by
\begin{eqnarray}
x \ = \ \alpha\left(p_x+ a_1y  \right)\, ,\;\;\;\;\; y \ = \
\alpha \left(p_y - a_1x    \right)\, ,
\end{eqnarray}
where $\alpha$ is an arbitrary real number. Note that $\alpha = 0$
and $\alpha = \infty$ also cover the singular cases $|{\bol{x}}| =
0$ and $|{\bol{A}}| = 0$, respectively. Inasmuch, instead of one
first-class condition $\phi = 0$ we can consider two second-class
constraints
\begin{eqnarray}
&&\phi_1 \ = \ \left( p_x - \frac{x}{\alpha} + a_1 y\right) \ = \
0\nonumber\\
&&\phi_2 \ = \ \left(p_y - \frac{y}{\alpha} - a_1x\right) \ = \
0\, ,
\end{eqnarray}
($\{\phi_1, \phi_2 \} = 2a_1 \neq 0$). Equivalently one may view
$\phi_1$ as a primary first-class constraint and $\phi_2$ as the
gauge fixing condition. To make contact with the Faddeejev-Jackiw
procedure we chose the second scenario. The corresponding reduced
Lagrangian is then
\begin{eqnarray}
&&\mbox{\hspace{-4mm}}L_R(\hat{{\bol{\xi}}},
\dot{\hat{\bol{\xi}}}) \  =  \ \dot{y}p_y +
\dot{x}\left(\frac{x}{\alpha} -a_1 y \right) - xp_y + y\left(
\frac{x}{\alpha} - a_1 y \right)\nonumber \\
&&~ \nonumber \\ &&\mbox{\hspace{10mm}} = \,
-\frac{1}{2a_1}\left(p_y + a_1 x - \frac{y}{\alpha}
\right)\frac{d}{dt}\left( p_x + a_1y - \frac{x}{\alpha}
\right)\nonumber \\
&&~ \nonumber \\ &&\mbox{\hspace{15mm}} - \, xp_y + y\left(
\frac{x}{\alpha} - a_1 y \right)\, . \label{6.8a}
\end{eqnarray}
At this point we can perform Darboux's transformation
\begin{eqnarray}
&&p_{\zeta} \ = \ \frac{1}{\sqrt{2}}\left(p_y +
a_1 x -
\frac{y}{\alpha} \right)\nonumber \\
&&\zeta \ = \ - \frac{1}{\sqrt{2} a_1}\left(p_x -\frac{x}{\alpha}
- a_1 y \right)\nonumber \\
&&z \ = \ \phi_2/2a_1 \ = \ \frac{1}{2a_2}\left(p_y -
\frac{y}{\alpha} - a_1x  \right)\, . \label{6.8}
\end{eqnarray}
%
%
The reduced Lagrangian (\ref{6.8a}) then becomes
\begin{eqnarray}
L_R({\bol{\zeta}}, \dot{\bol{\zeta}}, z)  =
\mbox{$\frac{1}{2}$}\zeta^s {\bol{\omega}_{st}}{\dot{\zeta}}^t -
\frac{1}{2a_1}p_{\zeta}^2 -  \frac{a_1}{2} \left(\zeta^2 - 2z^2
\right) , \label{6.6}
\end{eqnarray}
($\zeta \equiv \zeta^1$, $p_{\zeta} \equiv\zeta^2$). The
stabilization condition $\chi({\bol{\zeta}}, z) = 0$ in this case
yields the gauge fixing condition
\begin{eqnarray}
\chi({\bol{\zeta}}, z) \ = \ \frac{\partial H_R'({\bol{\zeta}},
z)}{\partial z} \ = \ - 2a_1 z  \ = \ 0\, .
\end{eqnarray}
By plugging $z = 0$ into Eq.(\ref{6.6}) (i.e., by enforcing a
gauge constraint) we eliminate the variable $z$ and obtain a
non-degenerate reduced Lagrangian
\begin{eqnarray}
L_R({\bol{\zeta}}, \dot{\bol{\zeta}})  \ = \
\mbox{$\frac{1}{2}$}\zeta^s {\bol{\omega}_{st}}{\dot{\zeta}}^t -
\frac{1}{2a_1}\ \!p_{\zeta}^2 -  \frac{a_1}{2}\ \!\zeta^2\, .
\end{eqnarray}
The canonical transformation ${\bol{\xi}} \mapsto ({\bol{\zeta}}, z,
p_z)$ is completed by identifying
\begin{eqnarray}
p_z \ = \  -\phi_1 \ = \ - p_x -a_1y + \frac{x}{\alpha}\, .
\label{6.9}
\end{eqnarray}
Note that, similarly as in the previous case, $\{p_\zeta, \zeta,
z, p_z\}$ can be identified with the Shouten-Eisenhart basis.

By choosing  $a_1 = 1/m\hbar$ and rescaling $\zeta(t) \mapsto
\zeta(t)/\hbar$ in the path integral (\ref{5.41}) we obtain the
quantum partition function for the linear harmonic oscillator with
a unit frequency. One can again observe that the fundamental scale
(suggestively denoted as $\hbar$) enters the partition function in
a correct quantum mechanical manner. This is precisely the result
which 't~Hooft conjectured for the system (\ref{6.1}) in
Ref.~\cite{tHooft3}.

Because the canonical transformation ${\bol{\xi}} \mapsto
({\bol{\zeta}}, z, p_z)$ is in this case linear it does not induce
anomaly in the path integral measure nor in the action (see also
Appendix~A).

%

In the framework of the Dirac-Bergmann treatment both results
discussed above were already derived in Ref.~\cite{BJK}. It is
clear that other emergent quantum systems can be generated in an
analogous manner. For instance, in Ref.\cite{BJK}
free particle weakly coupled to Duffing's oscillator was obtained
from the R\"{o}ssler system. Further development in this direction
is presently in progress.

\section{Summary}

Let us summarize the novel elements of this paper in comparison with
our previous work~\cite{BJK}. Here, we have utilized the
Faddeev-Jackiw treatment of singular Lagrangians~\cite{F-J} which
entirely obviates the need for the Dirac-Bergmann distinction
between first and second class, primary and secondary constraints
used in \cite{BJK}. Both approaches, however, require a doubling of
configuration space degrees of freedom. Apart from formulating the
path integral for singular Hamiltonians, the Faddeev-Jackiw method
is convenient in imposing 't~Hooft's information loss condition. In
the Dirac-Bergmann scheme, this condition represents a first-class
subsidiary constraint which has to be supplemented by a gauge fixing
condition~\cite{BJK}. In the Faddeev-Jackiw procedure the degrees of
freedom are reduced before quantization. This seems at first sight
simpler than the Dirac-Bergmann method, but it can be complicated in
practice. In particular, the change of coordinates (Darboux
coordinates) from the  pre-symplectic to a symplectic form plus
nondynamical ${\bol{z}}$-variables may be involved, or even
impossible. A detailed discussion of such difficulties can be found,
for instance, in Ref.~\cite{Pons}.

In the Dirac-Bergmann procedure, the reduction to physical degrees
of freedom is performed by dividing the constraints into first-class
and second class. Second-class constraints are removed via Dirac's
brackets machinery while the first-class constrains can be imposed
only after the gauge fixing procedure. In the Faddeev-Jackiw
treatment one does not need to classify constraints and perform
gauge fixing. Any possible gauge conditions are taken care of
implicitly by the reduction procedure. If the implicit gauge
conditions are global, it is possible to show~\cite{pons} that both
the Faddeev-Jackiw treatment and Dirac-Bergmann procedure leads to
the same reduced system. If they are only local, this equivalence
between the two schemes may be obstructed by unwanted Gribov
ambiguities. Thus, under the assumption that there exists a global
Darboux transformation we have shown that 't~Hooft's quantization
program performed with the Dirac-Bergmann and the Faddeev-Jackiw
procedure lead to  equivalent path integral representations of
emergent quantum systems.

Another problem may come from the the specific form of the Darboux
transformation. Although it is essentially non-canonical, it shows
up as a canonical transformation in the original configuration
space in which the constraints are embedded. Under normal
circumstances, the path integral measure is not
Liouville-invariant under canonical transformations, often
developing an anomaly~\cite{SW,EG}. This may invalidate our formal
path integral manipulations in Section~V. Fortunately, the
forewarned is also forearmed: if the canonical transformations are
linear it can be argued~\cite{SW} that an anomaly is not present.
This strategy seems to be simpler to utilize in the Dirac-Bergmann
than in Faddeev-Jackiw approach. This is because in the
Dirac-Bergmann analysis the gauge constraint is introduced by hand
(provided it is admissible) and hence one can try to choose it in
a way that the resulting canonical transformation is linear or at
least free of the anomaly.

In the Dirac-Bergmann approach it seems also easier to handle the
ordering problem mentioned in Section V. This is because
't~Hooft's constraint is there implemented directly via the linear
canonical transformation in the extended phase space. Due to the
fact that
\begin{eqnarray*}
\int_{t_1}^{t_2} dt \ ({\bol{p}} \dot{\bol{q}} -
H({\bol{p}},{\bol{q}})) = \int_{t_1}^{t_2} dt \ ({\bol{P}}
\dot{\bol{Q}} - H^*({\bol{P}},{\bol{Q}}))\, ,
\end{eqnarray*}
under a canonical transformation (modulo total derivative) there is
no explicit coordinate dependence in the term
$\dot{\bol{Q}}\bol{P}$. Thus the path integral is in this case well
defined even globally. This should be contrasted with the
Faddeev-Jackiw method where the phase space is not extended and
't~Hooft's constraint is imposed directly through a non-canonical
transformation.  Although the latter is only a halfway step toward
an ultimately canonical transformation it causes the path integral
to be well defined only perturbatively at these stages.

Note finally that according to analysis in Section~V, when we start
with the $N$--dimensional classical system (${\bol{q}}$ variables)
then the emergent quantum dynamics has $N-1$ dimensions
(${\bol{\zeta}}$ variables). This reduction of dimensionality
vindicates in part the terminology ``information loss" used
throughout the text.

\section*{Appendix A}

In the operator approach to quantum mechanical systems any
non-trivial change of variables is complicated by the ordering and
non-commutativity of the constituent operators that occur in
expressions. Due to $c$-number nature of path integrals such
difficulties are not immediately apparent. However, a careful
analysis of time-sliced representations of path integrals reveals
that complications related with canonical transformations are
hidden in two places~\cite{PI,SW,EG}. {\bf a)} the path-integral
phase space measure cannot be viewed as a product of Liouville
measures and, as a rule, canonical transformations often produce
anomaly
--- the Jacobian is not unity. {\bf b)} the time sliced canonical
transformation may generate in the action additional terms that
are of order $O(\Delta{\bol{P}})$ and $O(\Delta{\bol{Q}})$
(${\bol{P}}$ and ${\bol{Q}}$ are new variables, $\Delta X$ stands
for $X(t_{i+1}) - X(t_i)$), i.e., terms that need not vanish in
the continuous limit (i.e, when $\Delta t \equiv
\epsilon\rightarrow 0$). It is purpose of this appendix to show
that neither {\bf a)} nor {\bf b)} are hampering conclusions of
Section~VI.

As for {\bf a)}, it can be shown~\cite{SW} that to the lowest
order the anomalous inverse Jacobian for our canonical
transformation $(x,y,p_x,p_y) \equiv {\bol{\xi}} \mapsto
({\bol{\zeta}}, z, p_z)$ can be written as
\begin{eqnarray*}
J^{-1}  =  \prod_{j=1}^N(1 + A^{\zeta}_j \Delta \zeta_j + A^z_j
\Delta z_j + B^{\zeta}_j \Delta p^{\zeta}_j + B^z_j \Delta
p^z_j)\, ,
\end{eqnarray*}
where $\lim N \rightarrow \infty$ is understood  and
\begin{eqnarray}
A^{\zeta}_j &=& \mbox{$\frac{1}{2}$} \frac{\partial^3 F_j}{\partial
p^b_j \partial p^c_j \partial \zeta_j } \frac{\partial
p_j^b}{\partial\zeta_j}\frac{\partial \zeta_j}{\partial p^c_j} +
\mbox{$\frac{1}{2}$} \frac{\partial^3 F_j}{\partial p^b_j \partial
p^c_j \partial z_j } \frac{\partial
p_j^b}{\partial\zeta_j}\frac{\partial z_j}{\partial p^c_j} \nonumber
\\
&& + \ \mbox{$\frac{1}{2}$} \frac{\partial^3 F_j}{\partial p^x_j
\partial\zeta_j \partial\zeta_j  } \frac{\partial \zeta_j}{\partial x_j}
+ \mbox{$\frac{1}{2}$} \frac{\partial^3 F_j}{\partial p^x_j
\partial z_j \partial\zeta_j  } \frac{\partial \zeta_j}{\partial
y_j}\nonumber \\
&& + \ \mbox{$\frac{1}{2}$} \frac{\partial^3 F_j}{\partial p^y_j
\partial\zeta_j \partial\zeta_j  } \frac{\partial z_j}{\partial x_j}
+ \mbox{$\frac{1}{2}$} \frac{\partial^3 F_j}{\partial p^y_j
\partial z_j \partial\zeta_j  } \frac{\partial z_j}{\partial
y_j}\, ,\nonumber \\
A^{z}_j &=& \mbox{$\frac{1}{2}$} \frac{\partial^3 F_j}{\partial
p^b_j \partial p^c_j \partial \zeta_j } \frac{\partial
p_j^b}{\partial z_j}\frac{\partial \zeta_j}{\partial p^c_j} +
\mbox{$\frac{1}{2}$} \frac{\partial^3 F_j}{\partial p^b_j \partial
p^c_j \partial z_j } \frac{\partial p_j^b}{\partial
z_j}\frac{\partial z_j}{\partial p^c_j} \nonumber
\\
&& + \ \mbox{$\frac{1}{2}$} \frac{\partial^3 F_j}{\partial p^x_j
\partial\zeta_j \partial z_j  } \frac{\partial \zeta_j}{\partial x_j}
+ \mbox{$\frac{1}{2}$} \frac{\partial^3 F_j}{\partial p^x_j
\partial z_j \partial z_j  } \frac{\partial \zeta_j}{\partial
y_j}\nonumber \\
&& + \ \mbox{$\frac{1}{2}$} \frac{\partial^3 F_j}{\partial p^y_j
\partial\zeta_j \partial z_j  } \frac{\partial z_j}{\partial x_j}
+ \mbox{$\frac{1}{2}$} \frac{\partial^3 F_j}{\partial p^y_j
\partial z_j \partial z_j  } \frac{\partial z_j}{\partial
y_j}\, , \nonumber \\
B^{\zeta}_j &=& \mbox{$\frac{1}{2}$} \frac{\partial^3 F_j}{\partial
p_j^b \partial p_j^c \partial \zeta_j} \frac{\partial
p_j^b}{\partial p^{\zeta}_j} \frac{\partial \zeta_j}{\partial q_j^c}
+ \mbox{$\frac{1}{2}$} \frac{\partial^3 F_j}{\partial p_j^b \partial
p_j^c \partial z_j} \frac{\partial p_j^b}{\partial p^{\zeta}_j}
\frac{\partial z_j}{\partial q_j^c}\, , \nonumber \\
B^{z}_j &=& \mbox{$\frac{1}{2}$} \frac{\partial^3 F_j}{\partial
p_j^b \partial p_j^c \partial \zeta_j} \frac{\partial
p_j^b}{\partial p^{z}_j} \frac{\partial \zeta_j}{\partial q_j^c} +
\mbox{$\frac{1}{2}$} \frac{\partial^3 F_j}{\partial p_j^b \partial
p_j^c \partial z_j} \frac{\partial p_j^b}{\partial p^{z}_j}
\frac{\partial z_j}{\partial q_j^c}\, . \nonumber \\
\label{A1}
\end{eqnarray}
Here $F_j$ represents the classical generating function of the third
kind $F(p_x,p_y, \zeta, z)$ at the sliced time $t_j$. The new
variables are determined by solving the system of equations
\begin{eqnarray}
&&\mbox{\hspace{-8mm}}x = - \frac{\partial F(p_x,p_y, \zeta,
z)}{\partial p_x}, \;\; y = -
\frac{\partial F(p_x,p_y, \zeta, z)}{\partial p_y},\nonumber \\
&&\mbox{\hspace{-8mm}}p^{\zeta} = - \frac{\partial F(p_x,p_y,
\zeta,z)}{\partial \zeta}, \;\; p^z = - \frac{\partial F(p_x,p_y,
\zeta,z)}{\partial z}. \label{A2}
\end{eqnarray}
Indices $b,c$ in (\ref{A1}) run from $1$ to $2$ and summation
convention is assumed. It should be stressed that the higher order
contributions to the inverse Jacobian involve third and higher
order derivatives of $F(p_x,p_y, \zeta, z)$.

Straightforward but tedious calculations reveal that for the
canonical transformation (\ref{VI51}), (\ref{VI52}) we obtain
\begin{eqnarray}
A^z_j &=& \frac{\partial^3 F_j}{\partial p^x_j
\partial z_j \partial z_j  } \frac{\partial \zeta_j}{\partial
y_j} + \frac{\partial^3 F_j}{\partial p^y_j
\partial z_j \partial z_j  } \frac{\partial z_j}{\partial
y_j}\nonumber \\[2mm]
&=&
    - \left[ (1 + 2 a_1\,z \,p_{\zeta} ) \, \cos z +
      \left( p_z p_{\zeta}^{-1} - a_1p_{\zeta} \right) \,\sin z
      \right]
     \nonumber \\
&\times&   \left.\frac{\sin z}{2}\, \right|_{t=t_j}\, ,\nonumber \\[2mm]
A^{\zeta}_j &=& \ B^{\zeta}_j \ = \ B^z_j \ = \ 0\, .
\end{eqnarray}
The non-trivial contribution from $A^z_j$ is however zero at the
physical subspace because of the presence of $\delta[z]$
functional in the path integral measure of (\ref{V41}).

To complete the proof we must show that the leading-order form for
$J^{-1}$ is sufficient and that there is no need to go to higher
orders. This can be seen, for instance, from the exponentiated
form of the Jacobian:
\begin{eqnarray}
J^{-1} &=& \ e^{\sum_{j=1}^N\ln(1 + A^{\zeta}_j \Delta \zeta_j +
A^z_j \Delta z_j + B^{\zeta}_j \Delta p^{\zeta}_j + B^z_j \Delta
p^z_j)}
\nonumber \\[2mm]
&\approx& \ e^{\sum_{j=1}^N( A^{\zeta}_j \Delta \zeta_j + A^z_j
\Delta z_j + B^{\zeta}_j \Delta p^{\zeta}_j + B^z_j \Delta
p^z_j)}\, .
\end{eqnarray}
From the $\delta$-functionals in the measure we immediately have
that $\Delta z_j = \Delta p^{z}_j = 0$. On the other hand, from
the cyclicity of the Hamiltonian in $\zeta$ follows~\cite{SW} that
$\Delta p^{\zeta}_j = 0$ and $\Delta \zeta_j = O(\epsilon)$, i.e.,
the H\"{o}lder continuity index is $1$ rather than $1/2$ . So,
although there exists a contribution that can potentially bestow a
finite quantity on the action, namely
\begin{eqnarray*}
\exp\left[\sum_{j=1}^N A^{\zeta}_j \Delta \zeta_j  \right]
\rightarrow \exp{\left[\int_{t_i}^{t_f}dt \ A^{\zeta} \
\dot{\zeta} \right]}\, ,
\end{eqnarray*}
this term is trivial because $A_{j}^{\zeta} = 0$ for all $j$.

Similar analysis can be done for the canonical transformation
(\ref{6.8}), (\ref{6.9}). Since the transformation is linear,
$F(p_x,p_y, \zeta, z)$ must be quadratic and hence (\ref{A1})
implies that
\begin{eqnarray}
A^{\zeta}_j = A^z_j = B^{\zeta}_j = B^z_j = 0\, .
\end{eqnarray}
In this case the H\"{o}lder continuity index is $1/2$ as usual. So
by taking into account the constraints we have $\Delta p^z_j =
\Delta z_j =0$ and $\Delta p^{\zeta}_j = \Delta \zeta_j =
O(\sqrt{\epsilon})$. In general case we would need to consider also
terms of order $O(\epsilon)$ since the original Hamiltonian also
carries a factor of $\epsilon$ in the action (for our system are
higher orders in $\epsilon$ clearly irrelevant and can be omitted).
Fortunately, as already mentioned, higher order terms in $J^{-1}$
come from third (and higher) derivatives of $F(p_x,p_y, \zeta, z)$
and hence are identically zero for any linear canonical
transformation. Inasmuch the transformation (\ref{6.8}), (\ref{6.9})
does not produce any Liouville anomaly.

As for {\bf b)}, the situation is simpler in the case of a
transforation (\ref{VI51}), (\ref{VI52}). This is because the
transformation yields the Hamiltonian that is cyclic in $\zeta$
and $z$ which by itself ensures~\cite{SW} that any potential
pieces generated in the canonical transformation due to a finite
time slicing are of order $O(\epsilon^2)$ and hence disappear in
the path integral in the continuous limit.

In the case of transformation (\ref{6.8}), (\ref{6.9}) the
generating function reads
\begin{eqnarray*}
&&{\mbox{\hspace{-6mm}}}F(p_x,p_y, \zeta, z) \nonumber \\
&&= \frac{1}{2(\,a_1^2\,{\alpha }^2 -1)}\left[p_x^2\,\alpha +
p_y^2\,\alpha + 2\,a_1\,p_x\,\alpha \,\left( 2\,a_1\,z\,\alpha
\right.\right.
\nonumber \\[2mm] && - \left.
p_y\,\alpha  + {\sqrt{2}}\,\zeta \right) -
    2\,a_1\,p_y\,\alpha \,\left( 2\,z +
    {\sqrt{2}}\,a_1\,\alpha \,\zeta  \right)\nonumber \\[2mm]
&&+ \
    2\,a_1\,\left( {\sqrt{2}}\,z\,\zeta  + {\sqrt{2}}\,a_1^2\,z\,{\alpha }^2\,\zeta +
      \left. a_1\,\alpha \,\left( 2\,z^2 + {\zeta }^2 \right)
\right)\right]\, .\nonumber \\
\end{eqnarray*}
The new momenta and coordinates then fulfil symmetrized
equations~\cite{SW}
\begin{eqnarray}
p^{z}_j  &=& \frac{a_1}{(1- \,a_1^2\,{\alpha }^2)}\left[
2a_1\alpha^2 \ \!p_x - 2\alpha  \ \! p_y + 4a_1 \alpha \ \! z
\right.\nonumber \\
&+& \left. \left(
1 + a_1^2 \alpha^2 \right) \sqrt{2} \ \! \zeta\right]\nonumber \\
&-& \frac{a_1}{2(1- \,a_1^2\,{\alpha }^2)}\left[ 4 a_1 \alpha \ \!
\Delta z +  \sqrt{2}\left( 1 + a_1^2 \alpha^2 \right)\ \!
\Delta\zeta \right] \, ,\nonumber\\[2mm]
p^{\zeta}_j  &=& \frac{\sqrt{2} a_1}{(1- \,a_1^2\,{\alpha
}^2)}\left[\alpha \ \! p_x - a_1 \alpha^2 p_y + \left( 1 + a_1^2
\alpha^2 \right)z \right. \nonumber\\
&+& \left. \sqrt{2} a_1 \alpha \ \!\zeta \right]\nonumber
\\
&-& \frac{a_1}{\sqrt{2}(1- \,a_1^2\,{\alpha }^2)}\left[\left(1 +
a_1^2 \alpha^2 \right)\Delta z + \sqrt{2} a_1 \alpha \ \!
\Delta\zeta \right]\, ,
\nonumber \\[2mm]
x_j &=& \frac{\alpha}{(1- \,a_1^2\,{\alpha }^2)}\left[p_x + a_1
\left(2a_1 \alpha \ \!z - \alpha \ \!p_y + \sqrt{2}\ \!
\zeta\right)\right]
\nonumber \\
&+&  \frac{\alpha}{2(1- \,a_1^2\,{\alpha }^2)}\left[a_1 \alpha
\ \!\Delta p_y - \Delta p_x\right]\, , \nonumber \\[2mm]
y_j &=& \frac{\alpha}{(1- \,a_1^2\,{\alpha }^2)}\left[ p_y - a_1
\alpha \ \! p_x - a_1 \left( 2 \ \! z + \sqrt{2} a_1 \alpha
\ \! \zeta \right)\right]\nonumber \\
&+& \frac{\alpha}{2(1- \,a_1^2\,{\alpha }^2)}\left[ a_1 \alpha \ \!
\Delta p_x - \Delta p_y\right]\, . \label{A3}
\end{eqnarray}
Relations (\ref{A3}) yield $p_x$ and $p_y$ in terms of the new
variables. We can now utilize the leading order Taylor expansions
\begin{eqnarray*}
&&\mbox{\hspace{-4mm}}\Delta p_x \ = \ -\mbox{$\frac{1}{2}$} \
\!\Delta p^z + \frac{1}{\sqrt{2}a_1 \alpha} \ \! \Delta p^{\zeta} -
\frac{1}{\alpha}
\ \! \Delta z - \frac{a_1}{\sqrt{2}} \ \! \Delta \zeta\, , \\[2mm]
&&\mbox{\hspace{-4mm}}\Delta p_y \ = \ -\frac{1}{2a_1 \alpha} \
\!\Delta p^z + \frac{1}{\sqrt{2}}\ \! \Delta p^{\zeta} + a_1 \ \!
\Delta z + \frac{1}{\sqrt{2} \alpha}\ \! \Delta \zeta\, ,
\end{eqnarray*}
and substitute (\ref{A3}) into the old Hamiltonian. After imposing
the constraints $z_j = \Delta z_j = p^z_j = \Delta p^z_j = 0$ we
obtain
\begin{eqnarray}
&&\mbox{\hspace{-10mm}}(xp_y - yp_x)_j \nonumber
\\
&&\mbox{\hspace{-7mm}}\rightarrow \ \frac{1}{2a_1}\
\!(p^{\zeta}_j)^2 + \frac{a_1}{2}\ \!\zeta^2_j - \frac{1}{4a_1
\alpha}(\alpha \ \!
p^{\zeta}_j + \zeta_j) \Delta p^{\zeta}_j\nonumber \\[1mm]
&&\mbox{\hspace{-5mm}} + \left( \frac{a_1}{4} \ \! \zeta_j -
\frac{1}{4a_1 \alpha} \frac{(1 + 7 a_1^2 \alpha^2)}{(a_1^2 \alpha^2
-1)} p^{\zeta}_j \right)\Delta \zeta_j +  O(\epsilon)\, .
\end{eqnarray}
Because $\Delta p^{\zeta}_j = \Delta \zeta_j = O(\sqrt{\epsilon})$,
the contribution of the correction terms to the action is of order
$O(\epsilon^{3/2})$ which means that such terms are suppressed in
the continuous limit.


\section*{Acknowledgments}

The authors acknowledge an instigating communication with R.~Jackiw
and very helpful discussions with E.~Gozzi, J.M.~Pons, and
F.~Scardigli. P.J. was financed by of the Ministry of Education of
the Czech Republic under the research plan MSM210000018. M.B. thanks
MURST, INFN, INFM for financial support. All of us acknowledge
partial support from the ESF Program COSLAB.

\bibliography{apssamp}

\end{document}